\documentclass[11pt]{article}
\usepackage{epsfig} 
\newcommand{\half}{\mbox{\small{$\frac{1}{2}$}}} 
 
\newcommand{\Nf}{N_{\!f}} 
\newcommand{\MSbar}{\overline{\mbox{MS}}}

\begin{document}
\title{Three loop $\MSbar$ transversity operator anomalous dimensions for
fixed moment $n$ $\leq$ $8$}
\author{J.A. Gracey, \\ Theoretical Physics Division, \\ 
Department of Mathematical Sciences, \\ University of Liverpool, \\ P.O. Box 
147, \\ Liverpool, \\ L69 3BX, \\ United Kingdom.} 
\date{} 
\maketitle 
\vspace{5cm} 
\noindent 
{\bf Abstract.} We compute the anomalous dimensions of the transversity
operator at three loops in the $\MSbar$ scheme for fixed moment $n$ where
$n$~$\leq$~$8$. The results for the RI$^\prime$ renormalization scheme are
also provided for an arbitrary linear covariant gauge for $n$~$\leq$~$7$.

\vspace{-16cm}
\hspace{13.5cm}
{\bf LTH 730}

\newpage
Recently, the three loop anomalous dimension for the unpolarized twist-$2$ 
flavour non-singlet and singlet Wilson operators have been determined
analytically for all values of the operator moment $n$, \cite{1,2,3,4}. This 
was a formidable undertaking spanning ten years and relied on cutting edge
computer algebra and symbolic manipulation techniques implemented on high
performance computers. With the obvious necessity of such results to ensure
the complete two loop evolution of the Wilson coefficients and obtain more
precise estimates of quantities such as $\alpha_s(M_Z)$, there is also the
need to extend such computations to problems involving spin. Indeed in this
area, one quantity which will be of interest is that relating to transversity
which was originally introduced in \cite{5,6,7}. This corresponds to the 
probability of finding a quark in a transversely polarized nucleon polarized 
parallel to the nucleon versus that of the nucleon in the antiparallel 
polarization. Although experimentally it is harder to extract information on 
the transversity compared with usual deep inelastic scattering, one will still 
require the anomalous dimensions of the underlying operator to as high a loop 
order as is calculationally feasible for accurate renormalization group 
evolution. Currently the two loop anomalous dimensions are available for 
arbitrary moment, $n$, in the $\MSbar$ scheme, \cite{8,9,10,11,12}. However, 
given the symbolic manipulation machinery now available \cite{1,2,3,4,13}, it 
is clearly only a matter of time before the full $n$-dependent three loop 
$\MSbar$ results are determined. Although the underlying transversity operator 
differs from that of the Wilson operators of \cite{1,2,3,4}, calculationally it
is on a par with the non-singlet work of \cite{14}. Prior to the full 
$n$-dependent results of \cite{1,2,3,4} for the twist-$2$ Wilson operators, one
approach was to carry out a fixed moment determination of the anomalous 
dimensions. Essentially the first even moments to $n$~$=$~$16$ (apart from 
$n$~$=$~$14$) were determined, \cite{14,15,16}. Although eventually superceded 
by the analytic result, given the huge number of Feynman diagrams to evaluate 
by tedious recurrence relations, having information on the final results from 
an {\em independent} calculation provided an important crosscheck on the full 
$n$-dependent expressions. This is particularly the case when computations of 
gauge invariant quantities are simplified by choosing to work in the Feynman 
gauge. This substantially reduces the large number of integration by parts and 
hence the computation time is significantly smaller than, say, in an arbitary 
linear covariant gauge. In other words working in a specific gauge means that 
the internal strong check of observing the gauge parameter cancellation for a 
gauge independent quantity is crucially absent.

Given the potential determination of analytic anomalous dimensions for the
transversity operator in the foreseeable future, it is the purpose of this
article to repeat the approach of \cite{14} and provide the $\MSbar$ anomalous 
dimensions for fixed moments at three loops up to and including moment 
$n$~$=$~$8$. This builds on the low moment results of \cite{17,18} where the 
anomalous dimensions of the tensor current and second moment were determined. 
More recently the results for moments $n$~$=$~$3$ and $4$ were provided in 
\cite{19}. In \cite{18,19} the {\em primary} aim was to provide the finite 
parts of a specific Green's function in order to aid lattice measurements of 
the same quantity. In particular those measurements had to match onto the 
ultraviolet part of the Green's function and the provision of the answer at 
three loops was necessary to help make the extraction of lattice results as 
precise as possible. However, the work also required performing the 
renormalization in the lattice renormalization scheme, known as the modified 
regularization invariant (RI$^\prime$) scheme, \cite{20,21}. It has a continuum
definition which is discussed at length in \cite{22,17}. As the anomalous 
dimension of a gauge invariant operator is gauge dependent in a mass dependent 
scheme, the computation had to be performed in an arbitrary linear covariant 
gauge. Though for practical reasons the lattice calculations were performed in 
the Landau gauge. Therefore, as a second thread to this article, we will also 
determine the transversity anomalous dimensions in the RI$^\prime$ scheme in an
arbitrary linear covariant gauge. Though for computational reasons this will be
restricted to $n$~$\leq$~$7$.  

First, we discuss the basic properties of the transversity operator and outline
our computational strategy. The operator is defined by, \cite{8,9,10}, 
\begin{equation} 
{\cal O}^{\mu\nu_1\ldots\nu_n} ~=~ {\cal S} \bar{\psi} \sigma^{\mu\nu_1} 
D^{\nu_2} \ldots D^{\nu_n} \psi 
\label{opdef}
\end{equation} 
where $\sigma^{\mu\nu}$~$=$~$\half [ \gamma^\mu, \gamma^\nu ]$ and $D_\mu$ is
the covariant derivative involving the coupling constant $g$. The operation 
${\cal S}$ denotes symmetrization of the Lorentz indices 
$\{\nu_1,\ldots,\nu_n\}$ as well as ensuring that the operator is traceless 
according to the rules
\begin{eqnarray}
\eta_{\mu\nu_i}{\cal O}^{\mu\nu_1\ldots\nu_i\ldots\nu_n} &=& 0 ~~~~ 
(i ~\geq~ 2) \nonumber \\
\eta_{\nu_i\nu_j}{\cal O}^{\mu\nu_1\ldots\nu_i\ldots\nu_j\ldots\nu_n} &=& 
0 ~.  
\label{tracedef}
\end{eqnarray} 
To renormalize (\ref{opdef}) we follow a procedure similar to \cite{17,18}
where the operator is inserted at zero momentum into a quark $2$-point
function, $\langle \psi(p) {\cal O}(0) \bar{\psi}(-p) \rangle$, where $p$ is
the momentum. In order to apply the {\sc Mincer} algorithm, \cite{23}, written
in the symbolic manipulation language {\sc Form}, \cite{24,25}, one needs to 
saturate the Lorentz indices with the appropriate tensor. This is because the
{\sc Mincer} formalism can only be applied to massless three loop Lorentz
scalar $2$-point Feynman integrals, \cite{23}. Here, since we are not 
interested in the {\em finite} part of this Green's function, we merely 
multiply it by a Lorentz tensor which has the same symmetry and tracelessness 
properties as the original operator itself. This leads to an immediate 
algebraic simplification. When the fully symmetrized and traceless operator is 
inserted into the Green's function, there is a part involving products of the 
tensor $\eta^{\mu\nu}$. These derive from ensuring the tensor is overall 
traceless. However, when these terms multiply the projection tensor such terms 
will give zero. Therefore, in constructing the Feynman rules for the operator
insertion for the current calculation, one needs only to consider the part of 
the operator which is independent of the $\eta^{\mu\nu}$ tensors. In other 
words, the object  $\bar{\psi} \sigma^{\mu(\nu_1} D^{\nu_2} \ldots D^{\nu_n)} 
\psi$. For high moments, this represents a huge reduction in work such as the 
tedious but automatic derivation of the full Feynman rules, which can involve a 
significantly large number of terms. Instead the main work is in the 
construction of the projection tensor. However, this is achieved computer 
algebraically by writing down the complete set of independent objects built 
from one $\sigma^{\mu\nu}$ tensor, together with the appropriate numbers of 
$\eta^{\mu\nu}$ tensors and momenta $p^\sigma$ such that the number of free 
Lorentz indices equates with that of the original operator. These independent 
tensors are then symmetrized automatically with respect to the indices 
$\{\nu_1,\ldots,\nu_n\}$ and the arbitrary coefficients chosen so that the 
overall projection tensor is traceless according to (\ref{tracedef}). As was 
indicated in \cite{18,19}, there are three independent projections and 
therefore for the determination of the anomalous dimensions one needs only to 
select one of these for the projection procedure. It is preferable to choose 
the most algebraically compact one to minimize computation time. We note that 
this procedure is in contrast to the fixed moment strategy of \cite{14} where 
the null vector $\Delta_\mu$ was introduced. However, as was discussed in 
\cite{14,26}, one needs to handle the resulting integrals using a different 
projection technique. Finally, we note that our procedure automatically ensures
that there is no operator mixing, \cite{19}. The Feynman diagrams required for 
$\langle \psi(p) {\cal O}(0) \bar{\psi}(-p) \rangle$ are generated 
automatically with the {\sc Qgraf} package, \cite{27}, and converted to 
{\sc Form} input notation prior to the application of the three loop 
{\sc Mincer} algorithm. There are $3$ one loop, $37$ two loop and $684$ three 
loop Feynman diagrams and throughout we have used dimensional regularization in
$d$~$=$~$4$~$-$~$2\epsilon$ dimensions. 

We now record that the three loop $\MSbar$ transversity anomalous dimensions,
$\gamma^{(n)}(a)$, for $n$~$=$~$5$, $6$ and $7$ are  
\begin{eqnarray}  
\gamma^{(5)}(a) &=& \frac{92}{15} C_F a ~+~ \left[ 189515 C_A 
- 41674 C_F - 79810 T_F \Nf \right] \frac{C_F a^2}{6750} \nonumber \\ 
&& +~ \left[ \left( 190836000 \zeta(3) + 1975309075 \right) C_A^2 \right.
\nonumber \\
&& \left. ~~~~~-~ \left( 572508000 \zeta(3) + 325464235 \right) C_A C_F \right.
\nonumber \\
&& \left. ~~~~~-~ \left( 1192320000 \zeta(3) + 511395100 \right) C_A T_F \Nf
\right. \nonumber \\
&& \left. ~~~~~+~ \left( 381672000 \zeta(3) - 254723696 \right) C_F^2 
\right. \nonumber \\
&& \left. ~~~~~+~ \left( 1192320000 \zeta(3) - 989903260 \right) C_F T_F \Nf 
\right. \nonumber \\
&& \left. ~~~~~-~ 83718800 T_F^2 \Nf^2 \right] \frac{C_F a^3}{12150000} ~+~ 
O(a^4) 
\label{tra5ms} 
\end{eqnarray}  
\begin{eqnarray}  
\gamma^{(6)}(a) &=& \frac{34}{5} C_F a ~+~ \left[ 204770 C_A 
- 42129 C_F - 88810 T_F \Nf \right] \frac{C_F a^2}{6750} \nonumber \\ 
&& +~ \left[ \left( 707616000 \zeta(3) + 7527909825 \right) C_A^2 
\right. \nonumber \\
&& \left. ~~~~~-~ \left( 2122848000 \zeta(3) + 1373507730 \right) C_A C_F 
\right. \nonumber \\
&& \left. ~~~~~-~ \left( 4626720000 \zeta(3) + 1841332000 \right) C_A T_F \Nf 
\right. \nonumber \\
&& \left. ~~~~~+~ \left( 1415232000 \zeta(3) - 684744816 \right) C_F^2 
\right. \nonumber \\
&& \left. ~~~~~+~ \left( 4626720000 \zeta(3) - 3910683210 \right) C_F T_F \Nf 
\right. \nonumber \\
&& \left. ~~~~~-~ 320975800 T_F^2 \Nf^2 \right]
\frac{C_F a^3}{42525000} ~+~ O(a^4) 
\label{tra6ms} 
\end{eqnarray}  
and
\begin{eqnarray}  
\gamma^{(7)}(a) &=& \frac{258}{35} C_F a ~+~ \left[ 75266555 C_A 
- 15484767 C_F - 33149830 T_F \Nf \right] \frac{C_F a^2}{2315250} \nonumber \\ 
&& +~ \left[ \left( 3517994592000 \zeta(3) + 38365845513450 \right) C_A^2 
\right. \nonumber \\
&& \left. ~~~~~-~ \left( 10553983776000 \zeta(3) + 5978407701105 \right)
C_A C_F \right. \nonumber \\
&& \left. ~~~~~-~ \left( 24084527040000 \zeta(3) + 9039144860900 \right) 
C_A T_F \Nf \right. \nonumber \\
&& \left. ~~~~~+~ \left( 7035989184000 \zeta(3) - 4192441946262 \right) C_F^2 
\right. \nonumber \\
&& \left. ~~~~~+~ \left( 24084527040000 \zeta(3) - 20698675427220 \right)
C_F T_F \Nf \right. \nonumber \\
&& \left. ~~~~~-~ 1651311191600 T_F^2 \Nf^2 \right]
\frac{C_F a^3}{204205050000} ~+~ O(a^4) 
\label{tra7ms} 
\end{eqnarray}  
where $a$~$=$~$g^2/(16\pi^2)$, $\zeta(n)$ is the Riemann zeta function, $\Nf$
is the number of quark flavours and the colour group Casimirs are given by
\begin{equation}
\mbox{Tr} \left( T^a T^b \right) ~=~ T_F \delta^{ab} ~~,~~
T^a T^a ~=~ C_F I ~~,~~ f^{acd} f^{bcd} ~=~ C_A \delta^{ab}
\end{equation}
and $T^a$ are the generators for the colour group with structure constants
$f^{abc}$. We have also calculated the $n$~$=$~$8$ moment but modified the 
method used to find (\ref{tra5ms}), (\ref{tra6ms}) and (\ref{tra7ms}). In order
to reduce the computation time we performed the calculation in the Feynman 
gauge. Whilst this significantly reduces the number of integration by parts 
needed to be carried out by the {\sc Mincer} algorithm, for example, the check 
of gauge invariance of the final result in the $\MSbar$ scheme is removed. 
However, as a partial check on the automatic construction of the $n$~$=$~$8$ 
Feynman rules and projection tensor, we have performed the two loop calculation
in an arbitrary linear covariant gauge and reconstructed the gauge independent 
result of \cite{8,9,10,11,12}. Thus, at three loops in the $\MSbar$ scheme we 
have  
\begin{eqnarray}  
\gamma^{(8)}(a) &=& \frac{551}{70} C_F a ~+~ \left[ 1270588235 C_A 
- 251839827 C_F - 568470280 T_F \Nf \right] \frac{C_F a^2}{37044000} 
\nonumber \\ 
&& +~ \left[ \left( 57548150352000 \zeta(3) + 651153115163775 \right) C_A^2 
\right. \nonumber \\
&& \left. ~~~~~-~ \left( 172644451056000 \zeta(3) + 110529966326535 \right)
C_A C_F \right. \nonumber \\
&& \left. ~~~~~-~ \left( 411490679040000 \zeta(3) + 148380288276500 \right) 
C_A T_F \Nf \right. \nonumber \\
&& \left. ~~~~~+~ \left( 115096300704000 \zeta(3) - 55777651074312 \right) 
C_F^2 \right. \nonumber \\
&& \left. ~~~~~+~ \left( 411490679040000 \zeta(3) - 356756758487220 \right)
C_F T_F \Nf \right. \nonumber \\
&& \left. ~~~~~-~ 27902636165600 T_F^2 \Nf^2 \right]
\frac{C_F a^3}{3267280800000} ~+~ O(a^4) ~. 
\label{tra8ms}
\end{eqnarray}

As further checks on our results, (\ref{tra5ms}), (\ref{tra6ms}), 
(\ref{tra7ms}) and (\ref{tra8ms}), we note that for $\MSbar$ all the two loop 
expressions agree with those of \cite{8,9,10,11,12}, when they are evaluated 
for specific $n$. Moreover, (\ref{tra5ms}), (\ref{tra6ms}) and (\ref{tra7ms}) 
are clearly independent of the linear covariant gauge fixing parameter, 
$\alpha$. Further, the coefficients of the leading order large $\Nf$ term at 
three loops, corresponding to the $C_F T_F^2 \Nf^2$ term, agrees with the 
analytic evaluation at $O(1/\Nf)$ for arbitrary $n$ given in \cite{18}. A final
check is provided by the fact that the triple and double poles in $\epsilon$ of
the resulting operator renormalization constant are reproduced precisely in
agreement with the renormalization group equation prediction for all four 
cases. This is important since we followed the procedure of \cite{28} for 
renormalizing operators in automatic symbolic manipulation programmes. For the 
specific case of the Lie group $SU(3)$ we have 
\begin{eqnarray} 
\left. \gamma^{(5)}(a) \right|^{SU(3)} &=& \frac{368}{45} a ~-~ 
2\left[ 119715 \Nf - 1538939 \right] \frac{a^2}{30375} \nonumber \\
&& -~ \left[ 188367300 \Nf^2 + 8942400000 \zeta(3) \Nf + 12843253410 \Nf 
\right. \nonumber \\
&& \left. ~~~~~-~ 954180000 \zeta(3) - 144207743479 \frac{}{} \right] 
\frac{a^3}{82012500} ~+~ O(a^4) 
\end{eqnarray}  
\begin{eqnarray} 
\left. \gamma^{(6)}(a) \right|^{SU(3)} &=& \frac{136}{15} a ~-~ 
2\left[ 44405 \Nf - 558138 \right] \frac{a^2}{10125} \nonumber \\
&& -~ \left[ 240731850 \Nf^2 + 11566800000 \zeta(3) \Nf + 16107360420 \Nf 
\right. \nonumber \\
&& \left. ~~~~~-~ 1179360000 \zeta(3) - 183119500163 \frac{}{} \right] 
\frac{a^3}{95681250} ~+~ O(a^4) 
\end{eqnarray}  
\begin{eqnarray} 
\left. \gamma^{(7)}(a) \right|^{SU(3)} &=& \frac{344}{35} a ~-~ 
2\left[ 16574915 \Nf - 205153309 \right] \frac{a^2}{3472875} \nonumber \\
&& -~ \left[ 206413898950 \Nf^2 + 10035219600000 \zeta(3) \Nf 
\right. \nonumber \\
&& \left. ~~~~~+~ 13678917121415 \Nf - 977220720000 \zeta(3) 
\right. \nonumber \\
&& \left. ~~~~~-~ 156962874344971 \frac{}{} \right] 
\frac{a^3}{76576893750} ~+~ O(a^4) 
\end{eqnarray}  
and
\begin{eqnarray} 
\left. \gamma^{(8)}(a) \right|^{SU(3)} &=& \frac{1102}{105} a ~-~ 
\left[ 284235140 \Nf - 3475978269 \right] \frac{a^2}{27783000} \nonumber \\
&& -~ \left[ 20926977124200 \Nf^2 + 1028726697600000 \zeta(3) \Nf 
\right. \nonumber \\
&& \left. ~~~~~+~ 1381224814218690 \Nf - 95913583920000 \zeta(3) 
\right. \nonumber \\
&& \left. ~~~~~-~ 15957293707773841 \frac{}{} \right] 
\frac{a^3}{7351381800000} ~+~ O(a^4) ~.
\end{eqnarray}
where $C_F$~$=$~$4/3$, $C_A$~$=$~$3$ and $T_F$~$=$~$1/2$. 

Finally, we note that the results analogous to (\ref{tra5ms}), (\ref{tra6ms})
and (\ref{tra7ms}) in the RI$^\prime$ scheme are 
\begin{eqnarray} 
\gamma^{(5)}_{\mbox{\footnotesize{RI$^\prime$}}}(a) &=& \frac{92}{15} C_F a
\nonumber \\
&& +~ \left[ \left( 30825 \alpha^2 + 92475 \alpha + 1740690 \right) C_A 
- 166696 C_F - 676560 T_F \Nf \right] \frac{C_F a^2}{27000} \nonumber \\ 
&& +~ \left[ \left( 194197500 \alpha^4 + 1854279000 \alpha^3 
- 583200000 \zeta(3) \alpha^2 + 8993896875 \alpha^2 \right. \right. 
\nonumber \\
&& \left. \left. ~~~~~~-~ 6026400000 \zeta(3) \alpha + 30074295375 \alpha 
- 37353312000 \zeta(3) \right. \right. \nonumber \\
&& \left. \left. ~~~~~~+~ 356401468700 \right) C_A^2 \right. \nonumber \\
&& \left. ~~~~~+~ \left( 91239750 \alpha^3 - 209956950 \alpha^2 
- 4997987400 \alpha + 1076976000 \zeta(3) \right. \right. \nonumber \\
&& \left. \left. ~~~~~~~~~~~~-~ 60979980560 \right) C_A C_F \right. 
\nonumber \\
&& \left. ~~~~~-~ \left( 1726200000 \alpha^2 - 1555200000 \zeta(3) \alpha 
+ 10041363000 \alpha + 17858880000 \zeta(3) \right. \right. \nonumber \\ 
&& \left. \left. ~~~~~~~~~~~~+~ 253330505600 \right) C_A T_F \Nf \right. 
\nonumber \\ 
&& \left. ~~~~~+~ \left( 1289803200 \alpha + 27164160000 \zeta(3) 
- 22363266560 \right) C_F T_F \Nf \right. \nonumber \\
&& \left. ~~~~~+~ \left( 10686816000 \zeta(3) - 7132263488 \right) C_F^2 
+ 41629683200 T_F^2 \Nf^2 \right] \frac{C_F a^3}{340200000} \nonumber \\
&& +~ O(a^4) 
\label{tra5ri}
\end{eqnarray}  
\begin{eqnarray} 
\gamma^{(6)}_{\mbox{\footnotesize{RI$^\prime$}}}(a) &=& \frac{34}{5} C_F a
\nonumber \\
&& +~ \left[ \left( 33075 \alpha^2 + 99225 \alpha + 1957910 \right) C_A 
- 168516 C_F - 769360 T_F \Nf \right] \frac{C_F a^2}{27000} \nonumber \\ 
&& +~ \left[ \left( 119070000 \alpha^4 + 1160028000 \alpha^3 
- 364500000 \zeta(3) \alpha^2 + 5621954625 \alpha^2 \right. \right. 
\nonumber \\
&& \left. \left. ~~~~~~-~ 3766500000 \zeta(3) \alpha + 18833983125 \alpha 
- 22639824000 \zeta(3) \right. \right. \nonumber \\
&& \left. \left. ~~~~~~+~ 231206558900 \right) C_A^2 \right. \nonumber \\
&& \left. ~~~~~+~ \left( 49437000 \alpha^3 - 176477400 \alpha^2 
- 3356146800 \alpha - 1150848000 \zeta(3) \right. \right. \nonumber \\
&& \left. \left. ~~~~~~~~~~~~-~ 40247461840 \right) C_A C_F \right. 
\nonumber \\
&& \left. ~~~~~-~ \left( 1058400000 \alpha^2 - 972000000 \zeta(3) \alpha 
+ 6258495000 \alpha + 11741760000 \zeta(3) \right. \right. \nonumber \\ 
&& \left. \left. ~~~~~~~~~~~~+~ 165502535600 \right) C_A T_F \Nf \right. 
\nonumber \\ 
&& \left. ~~~~~+~ \left( 866102400 \alpha + 18040320000 \zeta(3) 
- 14470193920 \right) C_F T_F \Nf \right. \nonumber \\
&& \left. ~~~~~+~ \left( 6469632000 \zeta(3) - 3130262016 \right) C_F^2 
+ 27562342400 T_F^2 \Nf^2 \right] \frac{C_F a^3}{194400000} \nonumber \\
&& +~ O(a^4) 
\label{tra6ri}
\end{eqnarray}  
and 
\begin{eqnarray} 
\gamma^{(7)}_{\mbox{\footnotesize{RI$^\prime$}}}(a) &=& \frac{258}{35} C_F a
\nonumber \\
&& +~ \left[ \left( 12006225 \alpha^2 + 36018675 \alpha + 739917710 \right) C_A 
\right. \nonumber \\
&& \left. ~~~-~ 61939068 C_F - 292181680 T_F \Nf \right] 
\frac{C_F a^2}{9261000} \nonumber \\ 
&& +~ \left[ \left( 1588423567500 \alpha^4 + 15732651627000 \alpha^3 
- 4900921200000 \zeta(3) \alpha^2 
\right. \right. \nonumber \\
&& \left. \left. ~~~~~~+~ 76165633191375 \alpha^2 
- 50642852400000 \zeta(3) \alpha + 255327460003875 \alpha 
\right. \right. \nonumber \\
&& \left. \left. ~~~~~~-~ 301268627856000 \zeta(3) 
+ 3230428196219700 \right) C_A^2 \right. \nonumber \\
&& \left. ~~~~~+~ \left( 598161044250 \alpha^3 - 2827913095350 \alpha^2 
- 47764761023700 \alpha 
\right. \right. \nonumber \\
&& \left. \left. ~~~~~~~~~~~~-~ 30807568512000 \zeta(3) - 558295101833000 
\right) C_A C_F \right. \nonumber \\
&& \left. ~~~~~-~ \left( 14119320600000 \alpha^2 
- 13069123200000 \zeta(3) \alpha + 84542124471000 \alpha 
\right. \right. \nonumber \\ 
&& \left. \left. ~~~~~~~~~~~~+~ 164110847040000 \zeta(3) 
+ 2326270368265200 \right) C_A T_F \Nf \right. 
\nonumber \\ 
&& \left. ~~~~~+~ \left( 12326389941600 \alpha + 254163329280000 \zeta(3) 
- 198131779069280 \right) C_F T_F \Nf \right. \nonumber \\
&& \left. ~~~~~+~ \left( 84431870208000 \zeta(3) - 50309303355144 \right) C_F^2 
\right. \nonumber \\
&& \left. ~~~~~+~ 391057633881600 T_F^2 \Nf^2 \right] 
\frac{C_F a^3}{2450460600000} ~+~ O(a^4) ~. 
\label{tra7ri}
\end{eqnarray}  
To summarize, this renormalization scheme is defined by ensuring that the
{\em finite} part of the renormalized Green's function $\langle \psi(p) 
{\cal O}(0) \bar{\psi}(-p) \rangle$, multiplied by the projector, is given 
purely by its tree value only, \cite{20,21,17,18,19}. Further, it is important 
to note that in (\ref{tra5ri}), (\ref{tra6ri}) and (\ref{tra7ri}) the variables
$a$ and $\alpha$ are to be regarded as RI$^\prime$ quantities. The relation to 
the $\MSbar$ variables are given to three loops in \cite{17} for an arbitrary
linear covariant gauge. Since our main motivation is to provide the $\MSbar$
anomalous dimensions, the $n$~$=$~$8$ moment in the RI$^\prime$ scheme is 
clearly not available since the corresponding $\MSbar$ computation was 
restricted to the Feynman gauge.  

We conclude by noting that we have provided higher moments of the transversity 
operator at three loops in both the $\MSbar$ and RI$^\prime$ schemes. Together 
with the earlier results of \cite{20,21,18,19}, there are now eight fixed 
moment $\MSbar$ anomalous dimensions available prior to an explicit 
$n$-dependent computation. This is similar to the situation with regard to the 
flavour non-singlet twist-$2$ operator where there were seven fixed moment 
anomalous dimensions available prior to the provision of the full $n$-dependent 
expression. Whilst it is still in principle possible to compute even higher 
moments using the method we have discussed here, we believe we have reached a
computational limit beyond which it is not viable to proceed. For instance, for
the three loop $n$~$=$~$8$ Feynman gauge calculation, the necessary Feynman 
rules took around $36$ hours to be generated electronically on a dual opteron
$64$ bit SMP machine ($2$ GHz), resulting in just under $5.5$~$\times$~$10^6$ 
terms. The former is an order of magnitude larger than the time required for 
the $n$~$=$~$7$ case. 

\vspace{1cm}
\noindent
{\bf Acknowledgements.} The author thanks Dr P.E.L. Rakow and Dr C. McNeile
for valuable discussions.


\begin{thebibliography}{99} 
\bibitem{1} S. Moch, J.A.M. Vermaseren \& A. Vogt, Nucl. Phys. {\bf B688}
(2004), 101. 
\bibitem{2} S. Moch, J.A.M. Vermaseren \& A. Vogt, Nucl. Phys. {\bf B691}
(2004), 129. 
\bibitem{3} S. Moch, J.A.M. Vermaseren \& A. Vogt, Nucl. Phys. Proc. Suppl.
{\bf 135} (2004), 137. 
\bibitem{4} S. Moch, J.A.M. Vermaseren \& A. Vogt, Phys. Lett. {\bf B606}
(2005), 123. 
\bibitem{5} J.P. Ralston \& D.E. Soper, Nucl. Phys. {\bf B152} (1979), 109. 
\bibitem{6} R.L. Jaffe \& X. Ji, Phys. Rev. Lett. {\bf 67} (1991), 552;  
Nucl. Phys. {\bf B375} (1992), 527. 
\bibitem{7} J.I. Cortes, B. Pire \& J.P. Ralston, Z. Phys. {\bf C55} (1992), 
409. 
\bibitem{8} F. Baldracchini, N.S. Craigie, V. Roberto \& M. Socolovsky,
Fortschr. Phys. {\bf 30} (1981), 505. 
\bibitem{9} X. Artru \& M. Mekhfi, Z. Phys. {\bf C45} (1990), 669. 
\bibitem{10} A. Hayashigaki, Y. Kanazawa \& Y. Koike, Phys. Rev. {\bf D56}
(1997), 7350.  
\bibitem{11} W. Vogelsang, Phys. Rev. {\bf D57} (1998), 1886.  
\bibitem{12} J. Bl\"{u}mlein, Eur. Phys. J. {\bf C20} (2001), 683. 
\bibitem{13} J.A.M. Vermaseren, Int. J. Mod. Phys. {\bf A14} (1999), 2037.
\bibitem{14} S.A. Larin, T. van Ritbergen \& J.A.M. Vermaseren, Nucl. Phys. 
{\bf B427} (1994), 41. 
\bibitem{15} A. R\'{e}tey \& J.A.M. Vermaseren, Nucl. Phys. {\bf B604} (2001),
281. 
\bibitem{16} J. Bl\"{u}mlein \& J.A.M. Vermaseren, Phys. Lett. {\bf B606} 
(2005), 130. 
\bibitem{17} J.A. Gracey, Nucl. Phys. {\bf B662} (2003), 247. 
\bibitem{18} J.A. Gracey, Nucl. Phys. {\bf B667} (2003), 242. 
\bibitem{19} J.A. Gracey, JHEP {\bf 0610} (2006), 040.
\bibitem{20} G. Martinelli, C. Pittori, C.T. Sachrajda, M. Testa \& A. 
Vladikas, Nucl. Phys. {\bf B445} (1995), 81. 
\bibitem{21} E. Franco \& V. Lubicz, Nucl. Phys. {\bf B531} (1998), 641
\bibitem{22} K.G. Chetyrkin \& A. R\'{e}tey, Nucl. Phys. {\bf B583} (2000), 3. 
\bibitem{23} S.G. Gorishny, S.A. Larin, L.R. Surguladze \& F.K. Tkachov,
Comput. Phys. Commun. {\bf 55} (1989), 381. 
\bibitem{24} S.A. Larin, F.V. Tkachov \& J.A.M. Vermaseren, ``The Form version
of Mincer'', NIKHEF-H-91-18. 
\bibitem{25} J.A.M. Vermaseren, math-ph/0010025. 
\bibitem{26} S.A. Larin, P. Nogueira, T. van Ritbergen \& J.A.M. Vermaseren, 
Nucl. Phys. {\bf B492} (1997), 338. 
\bibitem{27} P. Nogueira, J. Comput. Phys. {\bf 105} (1993), 279. 
\bibitem{28} S.A. Larin \& J.A.M. Vermaseren, Phys. Lett. {\bf B303} (1993), 
334. 
\end{thebibliography}
\end{document}